# Temperature and thickness evolution and epitaxial breakdown in highly-strained BiFeO$_3$ thin films

Anoop R. Damodaran,[1] Sungki Lee,[1] Karthik Jambunathan,[1] Scott MacClaren,[2] and Lane W. Martin[1,*]

[1] Department of Materials Science and Engineering and Materials Research Laboratory
University of Illinois, Urbana-Champaign, Urbana, IL 61801

[2] Center for Microanalysis of Materials, Materials Reserach Laboratory, University of Illinois, Urbana-Champaign, Urbana, IL 61801

*lwmartin@illinois.edu

**Abstract**

We present the temperature- and thickness-dependent structural and morphological evolution of strain induced transformations in highly-strained epitaxial BiFeO$_3$ films deposited on LaAlO$_3$ (001) substrates. Using high-resolution X-ray diffraction and temperature-dependent scanning-probe-based studies we observe a complex temperature- and thickness-dependent evolution of phases in this system. A thickness-dependent transformation from a single monoclinically distorted tetragonal-like phase to a complex mixed-phase structure in films with thicknesses up to ~200 nm is the consequence of a strain-induced spinodal instability in the BiFeO$_3$/LaAlO$_3$ system. Additionally, a breakdown of this strain-stabilized metastable mixed-phase structure to non-epitaxial microcrystals of the parent rhombohedral structure of BiFeO$_3$ is observed to occur at a critical thickness of ~300 nm. We further propose a mechanism for this abrupt breakdown that provides insight into the competing nature of the phases in this system.

## I. INTRODUCTION

BiFeO$_3$ is a room-temperature multiferroic perovskite exhibiting antiferromagnetism that is coupled with ferroelectric order.[1,2] At room temperature, bulk BiFeO$_3$ assumes a rhombohedrally-distorted perovskite structure with an *R3c* space group.[3] Only recently have researchers begun in earnest to analyze the structure of the new polymorphs observed in highly-strained BiFeO$_3$ films. Early theoretical[4,5] and experimental[6,7,8] studies suggested the possibility of a tetragonally-distorted phase (derived from a structure with *P4mm* symmetry, *a ~ 3.665 Å*, and *c ~ 4.655 Å*) with a large spontaneous polarization. Soon after, enhanced electromechanical strains as large as 4-5% had been demonstrated in so-called mixed-phase BiFeO$_3$ thin films (occuring at a critical strain level of ~4.5% compressive strain) that exhibit a strain-induced structural mixture in which several polymorphs coexist.[8] The enhanced electromechanical response in these materials has been attributed to the thickness-dependent development of this complex mixed-phase structure and the ability for this material to reversibly transform under applied electric fields between these various phases.[9,10] Since the studies, additional information has come forth about these highly-strained films including the observations that the so-called tetragonal-like phase is monoclinically-distorted[11,12,13,14] and that other intermediate phases are present and play an essential role in the mixed-phase structures.[9] Recent reports of structural, magnetic and ferromagnetic transformations[15,16,17] in highly strained BiFeO$_3$/LaAlO$_3$ heterostructures near room temperature holds promise for giant piezoelectric, magnetoelectric, and piezomagnetic responses.

Further insight into the nature of the thickness-dependent evolution of these highly strained BiFeO$_3$ films can be gained by investigating related work on the epitaxial growth of other metastable phases.[18] It has long been known that epitaxial thin film strain has a strong role to play in the evolution of thin film structure. Typically in a mismatched film-substrate situation, the film is coherently strained (referred to as a commensurate state) to some point where it becomes too costly to continue to accommodate all the strain in the film. At this point so-called discommensuration (or the formation of strain relieving defects) occurs driving the system into an incommensurate state. The mean separation distance between these strain-relieving defects generally decreases as the mismatch increases. Often these defects are misfit dislocations that form ordered arrays at the substrate/film interface.[19,20] The density of these misfit dislocations will increase as the film thickness is increased until the total strain in the film is reduced to zero and the lattice parameters return to those of the bulk. Following the nomenclature used by Bruinsma and Zangwill,[18] we will refer to coherent-incoherent transitions resulting from a variation in thickness ($h$) and commensurate-incommensurate transitions resulting from variations in lattice misfit ($f$).

It has been observed that in metal systems, where dislocation motion is relatively easy, predicted values of critical thicknesses ($h_c$) and thickness-dependence of coherency loss follow each other closely.[21] Oxide-based systems, however, are widely observed to deviate from these predictions due to large kinetic barriers to dislocation nucleation and migration.[22] Thus in these systems, alternative pathways for strain relaxation are necessary – including having the film adopt a crystal structure that is well lattice matched to the substrate, but that is different from the bulk structure of the film material. This process is referred to as *pseudomorphism* and the pseudomorphic phase is often coherently strained to the substrate. The study of such pseudomorphs dates back to the 1950s when alkali halide films were observed to undergo a pseudomorphic phase transformation.[23] Additionally, early molecular beam epitaxy studies found the even in certain metal systems, pseudomorphic phase transitions were possible. For instnace, work on Sb [which normally possesses a tetragonal BCC structure (white tin) with $a = 5.831$Å and $c = 3.181$Å at room temperature] found that this material adopted a low-temperature diamond structure (grey tin, $a = 6.489$Å) when deposited on (001) InSb and CdTe ($a = 6.48$Å).[24] By undergoing the pseudomorphic transformation, the Sb avoids an unfavorable lattice mistmatch and strain condition. Likewise, similar results have been obtained for Co films on GaAs.[25] More surprising in this case, films of Co <100 nm in thickness were found to grow as a previously unknown, metastable BCC version ($a = 2.819$Å) on GaAs (110) while films >100 nm were found to transform to the bulk HCP structure. More recently, Bruinsma and Zangwill[18] proposed a thickness dependent structural phase diagram as a function of the geometric misfit between the substrate and film and overall film thickness to help explain such effects. These predictions also include an intermediate strain regime where the film evolves from a single-phase highly-strained metastable structure to a spinodal-modulated mixed-phase structure before eventual breakdown to microcrystallites of the bulk stable phase. In the remainder of the manuscript we will investigate the applicability of this model to the observed features of the thickness dependent growth of highly-strained $BiFeO_3$ films on $LaAlO_3$ (001) substrates. We will establish a thorough understanding of the growth, thickness and temperature-dependant evolution of these highly-strained structures, their stability, and the role and influence of the parent rhombohedral-phase. The current work examines the evolution of these various phases, provides a proposed mechanism for the evolution of the mixed-phase structures important for the large electromechanical responses, exmines the eventual epitaxial breakdown of this system, and frames these results as a competition between the thermodynamically stable equilibrium rhombohedral-phase and the strain-induced polymorphs.

II. EXPERIMENTAL

Epitaxial BiFeO$_3$ films of thickness 20-400 nm were synthesized via pulsed laser deposition from Bi$_{1.1}$FeO$_x$ targets at 700°C in oxygen pressures of 100 mTorr on single-crystal LaAlO$_3$ (001) substrates and were cooled in oxygen pressures of 760 Torr. The laser fluence and repetition rate were maintained at 1.4 J/cm$^2$ and 10 Hz, respectively, for all growths resulting in an effective growth rate of ~0.28 Å/s. Care was taken to assure uniform deposition and appropriate chemistry and thus no single target was used to deposit more than 75 nm of material. Detailed structural information of the various films was obtained using high-resolution X-ray diffraction (X'Pert MRD Pro equipped with a PIXcel detector, Panalytical) including θ-2θ scans and reciprocal space maps (RSMs). Topographic studies of the as-grown films were carried out using temperature-dependent (25°C to 300°C) atomic force microscopy (AFM) (Cypher and MFP-3D, Asylum Research). The surface structure and cross-sections of the as-grown films were also observed using a Hitachi S-4800 high resolution Scanning Electron Microscope (SEM).

## III. RESULTS AND DISCUSSION

Typical θ-2θ X-ray diffraction studies about the 001-diffraction condition of BiFeO$_3$ films of thicknesses 30, 140, 250, and 350 nm [Fig. 1] reveal an interesting evolution in structure with thickness. Following the nomenclature established in recent studies,[9] the various phases observed are labeled as the rhombohedral parent phase (R-phase, $c$ = 3.96 Å), the intermediate monoclinic phase (M$_I$-phase, $c$ = 4.17 Å), and the monoclinically-distorted, tetragonal-like phase (M$_{II}$-phase, $c$ = 4.67Å). The 30 nm thick film exhibits a single peak corresponding to an out-of-plane lattice parameter ($c$) of ~4.67 Å, consistent with the M$_{II}$-phase. Upon increasing the film thickness, additional peaks corresponding first to the M$_I$-phase ($c$ = 4.17Å) and subsequently to the bulk-like R-phase ($c$ = 3.967Å) begin to appear. From our studies, we have observed that in films less than ~150 nm, the peak corresponding to the R-phase has very low intensity or is totally absent in some cases. By the time the thickness reaches ~250 nm, the presence of an R-phase peak is more noticeable for most films and by a thickness of ~350 nm only the peak corresponding to the R-phase is observed and all other peaks are completely absent. It should also be noted that the R-phase peak is considerably less intense than the peaks for the $M_{II}$-phase in the thinner films and, in general, shows lower diffraction intensities throughout the films studied. We also note that the out-of-plane lattice parameter of the $M_{II}$-phase increases from 4.63Å to 4.68Å as we transition from the 30 nm to 250 nm thick films [Fig. 1]. This suggests a rather complex thickness dependent evolution and strain relaxation process in these films.

Such observations present two important questions: what happens to the $M_{II}$-phase in thicker films and why does the R-phase peak intensity remain so low even in thick films? Here we develop a detailed picture of the complex behavior observed in these diffraction experiments and provide insight into the thickness-dependent evolution of this complex system. An understanding of the structural evolution is obtained by investigation of the surface topography of these films. We have investigated the surface topography of the films at various magnifications using both optical microscopy and AFM [Fig. 2]. Under the optical microscope, the 30 nm thick films are found to have an optically smooth surface (note the presence of the structural twins in the $LaAlO_3$ substrate visible in the image) [Fig. 2(a)], which is consistent with the AFM images [Fig. 2(b)] which exhibits only the $M_{II}$-phase with atomically smooth terraces, separated by single unit cell step-heights (~4.65Å). Likewise, the optical micrographs of the 140 nm thick films reveal these films to be optically smooth as well [Fig. 2(c)] and upon close inspection using AFM, we observe mixed-phase topography consisting of regions of atomically flat terraces of the $M_{II}$-phase [bright areas, Fig. 2(d)] and mixed-phase regions consisting of an intimate mixture of the $M_I$ and $M_{II,tilt}$-phases [striped regions, Fig. 2(d)], consistent with previous reports.[9]

Inspection of optical micrographs of the 250 nm thick films, on the other hand, reveal a surface that is mostly smooth with a number of rough regions [Fig. 2(e)]. We note that the fraction of these rough regions scales with thickness and does not appear to grow with additional time spent at high-temperatures without additional material being added to the surface. *Ex-situ* anneals at 500-600°C in oxygen for over 20 hours did not result in a change in the fraction of the rough regions. AFM studies of the optically flat regions [red box in Fig. 2(e), Fig 2(f)] once again reveal topography consistent with flat terraces of the $M_{II}$-phase and striped mixed-phase regions. We note that upon increasing the thickness from 140 nm to 250 nm the surface depressions associated with the mixed-phase regions increase greatly from ~7 nm to ~11 nm, respectively. Interestingly, however, AFM studies of the same sample in the rough regions [yellow box in Fig 2(e), Fig. 2(h)] show a significantly roughened surface with a peak-to-valley height scale of over 200 nm (nearly the entire thickness of the film) without any resemblance to the mixed-phase structures observed elsewhere on this sample. Further inspection of the 350 nm thick films under the optical microscope reveals that the rough regions have grown dramatically to cover the entire film surface [Fig. 2(g)]. Analysis of these films with AFM revealed surface morphologies similar to that observed in Fig. 2(h). The region within the black box in Fig. 2(h) is consistent with regions observed across this and other samples in this thickness range and evokes images of recent advances in the study of $BiFeO_3$ materials – especially the synthesis of $BiFeO_3$ single crystals via the flux growth method.[26,27,28]

Fig. 3(a) is a high-resolution AFM image of the area highlighted in Fig. 2(h) and reveals that the rough regions possess micron-sized crystallites with well-defined facets. These features bear a striking resemblance to BiFeO$_3$ single crystals grown by the flux method [Fig. 3(b)][26] which exhibit large flat (012) surfaces (using the crystallographic reference frame of the parent rhombohedral structure). Detailed high-resolution X-ray diffraction scans of our 350 nm samples have allowed us to obtain evidence for a number of peaks corresponding to the bulk-like R-phase of BiFeO$_3$ [Fig. 3(c)]. These diffraction patterns can be indexed by peaks corresponding to the most intense reflections from the diffraction patterns of BiFeO$_3$ single-crystals grown by the flux method [bottom, Fig. 3(b)]. We have even observed unique features of the bulk R-phase diffraction pattern such as the splitting of the 104- and 110- diffraction peaks in such films. This combination of X-ray diffraction and AFM strongly suggests that the rough, patchy regions are in fact regions of the bulk-like R-phase of BiFeO$_3$ that grow at the expense of the M$_I$- and M$_{II}$- phases in a non-epitaxial manner. We note that for each thickness reported here, we have included in the same growth a DyScO$_3$ (110) substrate for further analysis and comparison of the rhombohedral-like thin film phase. Similar inspection of the co-deposited BiFeO$_3$/DyScO$_3$ (110) films reveals smooth surfaces (both from optical microscopy and AFM, Suppl. Fig. S1) for all films up to and including the 350 nm thick films and show no evidence of second phases from X-ray diffraction.

We can further our understanding of the mechanism of strain accommodation and epitaxial breakdown in this system by analyzing the change in surface structure of a number of BiFeO$_3$ films with thickness ranging from 40 nm to 250 nm upon heating from room temperature to 300°C. Fig. 4 shows AFM topography images of films of three representative thicknesses 40 nm [Figs. 4(a)-(c)], 110 nm [Figs. 4(d)-(f)], and 250 nm [Figs. 4(g)-(i)] at three representative temperatures (moving left-to-right, 50°C, 200°C and 300°C). At any given temperature, the films reveal an increasing fraction of the mixed-phase regions with increasing film thickness (consistent with prior reports).[8] The reported fraction of the mixed-phase is calculated as the areal fraction of the mixed-phase regions relative to the entire area of the sample [Fig. 4(j)]. We also report the depth of the mixed-phase stripe-regions relative to the atomically flat plateau regions of the $M_{II}$-phase [trench depth, Fig. 4(k)] and the root-mean-square (RMS) roughness of these films which is an indicator of the volume fraction of the mixed-phase regions in these films [Fig. 4(l)]. Beginning with the thinnest film reported here (40 nm), we observe that ~20% of the areal fraction of the surface is made up of the mixed-phase regions and that this fraction decreases steadily to zero by 300°C, resulting in a terraced surface with unit cell step-heights corresponding to the $M_{II}$-phase [Fig. 4(a)-(c)]. We note that similar stripe-like mixed-phase regions are found to reappear upon cooling, but despite similarities in the location of features, they do not appear to have an exact memory for location and fine structure. Similar decreasing trends in the fraction of the mixed-phase are observed for both the 110 nm and 250 nm thick films; however, both of these films still exhibit a significant fraction of mixed-phases even at 300°C (the maximum we can achieve in our scanning probe system). Thus, we conclude that the temperature at which the film transforms to being composed entirely of the $M_{II}$-phase is a function of the film thickness and is higher for thicker films. This suggests that the films form the mixed-phase upon cooling down from the growth temperature and there exists a critical thickness at which the film will stabilize in the mixed-phase structure even at the growth temperature of 700°C.

As illustrated by the AFM experiments, these samples exhibit a temperature induced reduction in the fraction of the mixed-phase. We note that these results are consistent with the work in the supplementary materials of Ref. 8 where phase field simulations suggest a driving force for the stabilization of the highly-distorted $M_{II}$-phase with increasing temperature. We see that films up to a thickness of 35 nm grow as the $M_{II}$ phase which is stable down to room temperature. However, in thicker films (40-200 nm) we contend that the samples grow as a fully strain-stabilized, $M_{II}$-phase at 700°C and upon cooling, the mixed-phase structures are formed to accommodate the increase in strain energy. This suggests that the formation of the mixed-phase structure stabilizes the strained film at lower temperatures. It would appear that in this system, that instead of generating misfit dislocations in the sample, the material undergoes partial relaxation via the formation of the $M_I$- and $M_{II,tilt}$- mixed-phase regions. We also note that these mixed-phase stripe bands generally form 2D arrays on the sample of the surface with the long-axis of bands running along [100] and [010] in-plane directions. Such a configuration has parallels to classic 2D arrays of misfit dislocations.

We can better understand the nature of the formation of such mixed-phase structures during the cool-down process by investigating the energetics of the system. Figs. 5(a)-(c) show a schematic of the free energy landscape for films with thickness between 40-200 nm as a function of substrate induced strain $\varepsilon$ at different temperatures. Here we focus on a film with a thickness of 40 nm as an example. Theoretical calculations and experimental studies have suggested the presence of a number of different structural varieties of distorted $BiFeO_3$ with a range of $c/a$ lattice parameter ratios,[13,29,30] the most important of which for this discussion are the parent R-phase and the highly-distorted, strain induced $M_{II}$-phase. Thus the energy landscape should be characterized by at least two local minima corresponding to these two phases. At the growth temperature (700°C), we can thus draw a schematic energy diagram as a function of thin film strain such as that in Fig. 5(a). Consistent with previous experimental and density functional theory studies, growth at low strain levels (less than ~4% compressive strain) results in the formation of films possessing the R-phase structure while growth at strain levels in excess of 4% result in stabilization of the $M_{II}$-phase. Since the R-phase is the thermodynamically stable equilibrium phase at low-temperature and strain, the effect of cooling the film down from the growth temperature is to shift the energy minima for the strained metastable $M_{II}$-phase to higher energies and strains relative to the R-phase. Thus, as we cool the film from the growth temperature down to 300°C, the energy curves shift as noted. The region within the interval [$\varepsilon^-$, $\varepsilon^+$] with a negative curvature for the free energy forms a strain-induced spinodal and in this interval of substrate-induced strain, the film spontaneously splits to a modulated mixed-phase structure of alternating R- and $M_{II}$-like phases. The region of negative curvature shifts towards the strain condition for the film on the $LaAlO_3$ substrate upon cooling from the growth temperature [Fig. 5(b)]. Therefore, at room temperature the $LaAlO_3$ substrate forces the strain condition of the film into the strain-induced spinodal [Fig. 5(c)].[18,31] In this region, the film is mechanically unstable against local strain wave distortions and this drives a lowering of the energy by spontaneous deformation to the mixed-phase structures along the easy strain axes (<100>). Therefore, films exposed to these strain conditions, as a result of the interplay between thermal expansion mismatch, epitaxial strain, and thermodynamic phase stability, will spontaneously separate into a modulated mixed-phase structure of alternating R-like and $M_{II}$-like phases in the $BiFeO_3$ system.

The majority or our discussion thus far has focused on films with thickness less than 200-250 nm, but beyond this critical thickness we have observed epitaxial breakdown in these films. We now focus on the nature of this epitaxial breakdown. Fig 6(a) is a SEM cross-section of a 250 nm thick $BiFeO_3/LaAlO_3$ (001) films that was observed to have a small fraction of the rough regions reported in the optical micrographs [Fig 2(e)]. The presence of these rough regions marks the initial onset of epitaxial breakdown of the film. The SEM cross-section cuts across the optically smooth regions as well as the rough region [Fig. 6(a)]. A closer look at the cross-section of the optically smooth regions [blue box in Fig. 6(a), Fig. 6(b)] reveals a mixed-phase structure composed of alternating regions of $M_I$- and $M_{II,tilt}$-phases with sharp well-defined interfaces (emphasized by the yellow dotted lines). Focusing, in turn, on the interface between the smooth and rough regions [orange box in Fig. 6(a), Fig. 6(c)] we observe the formation of microcrystallites of the bulk R-phase (consistent with AFM and XRD studies) and that the breakdown, once initiated, is not limited to the surface but occurs through the entire thickness of the film. Note that the peak-to-valley roughness in these rough regions are found to be, in general, a good fraction of the entire film thickness. Fig. 6(d) is a cross-sectional image of a 350 nm thick film that reveals a complete breakdown of the film. Plan-view images [Fig. 6(e)] shows sharp faceted microcrystallites of the R-phase over the entire surface indicating a complete breakdown of epitaxy.

Based on these results, we can now begin to construct a structural phase diagram [Fig. 7(a)] at the deposition temperature of 700°C to help explain the evolution of the complex structure and morphology of these highly-strained $BiFeO_3$ films as a function of increasing film thickness. We note that this diagram is similar to the diagram proposed by Brunisma and Zangwill[18] for unrelated systems. The diagram shows the expected microstructure of the film as a function of epitaxial lattice mismatch between film and substrate and film thickness. Focusing first on the lattice misfit corresponding to the $LaAlO_3$ substrate, we note that for thickness <200 nm films grow in the pure $M_{II}$-phase and are coherently strained to the substrate [Fig. 7(b)]. The growth is expected to occur in a layer-by-layer or step-flow growth mode as the resulting $M_{II}$-phase regions reveal atomically flat terraces following growth. Note that films in excess of 35 nm will undergo a temperature-induced spinodal phase separation upon cooling. As the films with the strain-stabilized $M_{II}$-phase grow in thickness, so does the cost in free energy compared to the ground-state R-phase. At a critical thickness, energetics require that the films undergo a first order transformation to the bulk, stable crystal structure. However, large crystallographic deformations and geometric constraints associated with such a transformation present substantial kinetic barriers to the nucleation and transformation to the bulk, stable phase and this prevents the observation of the true equilibrium structure. As the film thickness approaches ~250 nm it enters the regime of high-temperature, thickness-driven, strain-relaxation-induced spinodal instability and forms a strain modulated structure of alternating $M_I$ and $M_{II,tilt}$ phases [Fig. 7(c)]. The spontaneous transformation to the mixed-phase structure is accompanied by surface relief with depressions that are easily several nanometers deep (roughly 4-5% of the film thickness) and results in the significant roughening of the growth front (i.e., the saw-tooth structure reported previously).[9] Several theoretical and experimental studies of systems undergoing spinodal phase separation and concomitant roughening of the growth front have demonstrated changes in growth mode resulting in film-to-island morphological transitions, including possible film break-up.[32,33] Moreover, such a mixed-phase structure with periodic interphase boundaries and surface relief significantly lowers the kinetic barriers to the nucleation of the bulk R-phase and as it approaches a thickness of ~ 300 nm, the film breaks down to non-epitaxial microcrystals of the bulk R-phase [Fig. 7(d)].

Furthermore, this phase diagram is consistent with previously observed work on $BiFeO_3$ thin films grown on other substrates. For instance, growth of $BiFeO_3$ on $YaAlO_3$ (110) substrates [$a$ = 3.71Å, large lattice mismatch, Fig. 7(a)] has been found to result in essentially phase-pure $M_{II}$-phase films up to thicknesses of 225-250 nm.[8] Likewise much work on $BiFeO_3$ thin films on $SrTiO_3$ (001) substrates [$a$ = 3.905Å, small lattice mismatch, Fig. 7(a)] has been reported and it has been observed that BiFeO$_3$ films will relax to incoherent films at thicknesses in excess of a few hundred nanometers.[13]

IV. CONCLUSIONS

These results have added to our understanding of these complex and technologically exciting phase boundaries in highly-strained BiFeO$_3$ thin films. The presence of a variety of polymorphs of the BiFeO$_3$ is essential for the strong electromechanical response in these films. We observe, however, that these structures are limited by a thickness-dependent breakdown and irreversible transformation to a non-epitaxial R-phase. We have examined the thickness- and temperature-dependence of these structures and have constructed schematic energy and phase diagrams to help explain the structural evolution of these materials. We have drawn parallels to observations of unusual strain-relaxation in more simplistic metallic systems and have applied a model for spinodally-modulated structures to BiFeO$_3$. The ability of the BiFeO$_3$ system to take on a variety of pseudomorphs provides one route to strain relaxation in this system and due to the complex interplay of lattice and electronic order in these materials this results in strong electromechanical responses. Our observations provide new insights into the nature of the phase evolution in highly compressively strained BiFeO$_3$, the stability of the various polymorphs, and are consistent with previously observed structures in a variety of epitaxial BiFeO$_3$ films. Equipped with such an understanding of the thickness-driven breakdown of epitaxy, we can begin to construct pathways to stabilize the desired mixed-phase structures in these exciting and technologically relevant materials.

The authors would like to acknowledge the help and scientific insights of Dr. S. MacLaren and Dr. M. Sardela at the Center for Microanalysis of Materials at UIUC. The work at UIUC was supported by the Army Research Office under grant W911NF-10-1-0482 and by Samsung Electronics Co., Ltd. under grant 919 Samsung 2010-06795. Experiments at UIUC were carried out in part in the Frederick Seitz Materials Research Laboratory Central Facilities, which are partially supported by the U.S. Department of Energy under grants DE-FG02-07ER46453 and DE-FG02-07ER46471.

**Figures**

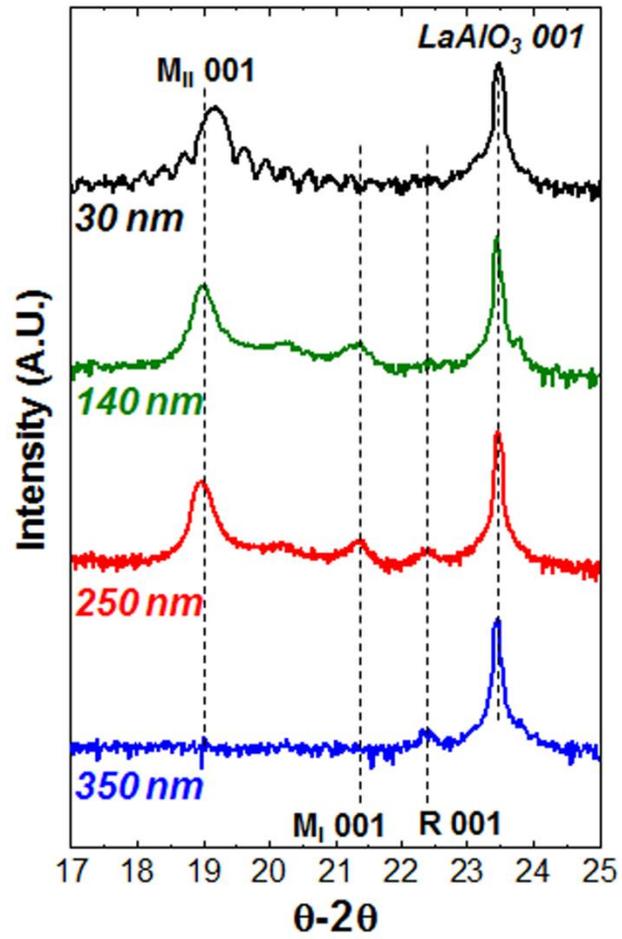

FIG. 1. X-ray diffraction about the 001 diffraction condition of $BiFeO_3/LaAlO_3$ (001) heterostructures for (top-to-bottom) 30 nm, 140 nm, 250 nm, and 350 nm thick films.

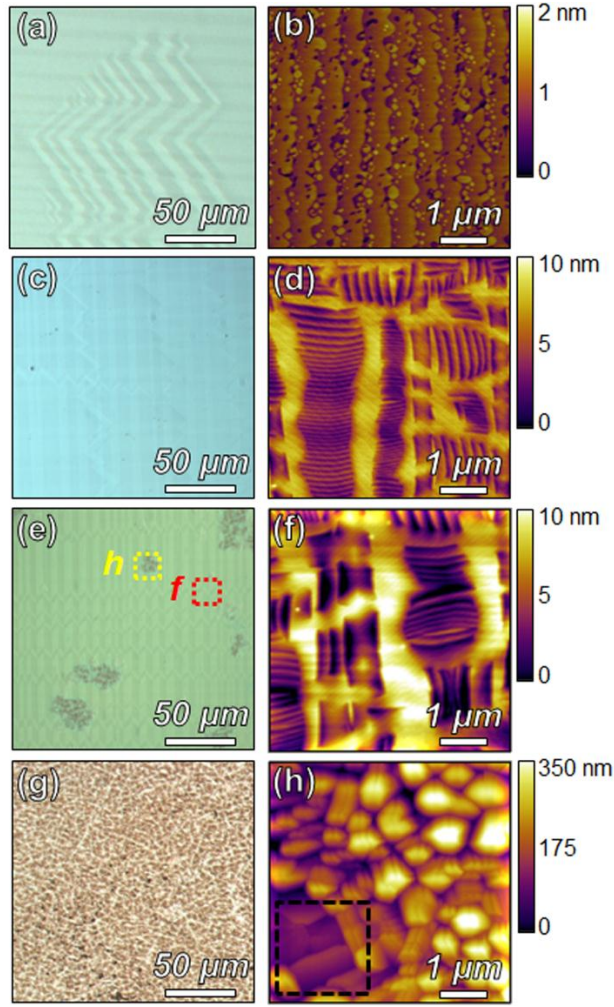

FIG. 2. Optical (left) and atomic force microscopy (right) images of BiFeO$_3$/LaAlO$_3$ (001) heterostructures of various thicknesses. (a) and (b) are images for a 30 nm thick film and (c) and (d) for a 140 nm thick film. (e) reveals formation of different types of structures in 250 nm thick films. Close inspection of (f) of the smooth areas reveals results consistent with thinner films and investigation of patchy regions reveals rough microstructure (h). (g) is an optical micrograph of a 350 nm thick film which is found to possess only the rough microstructure.

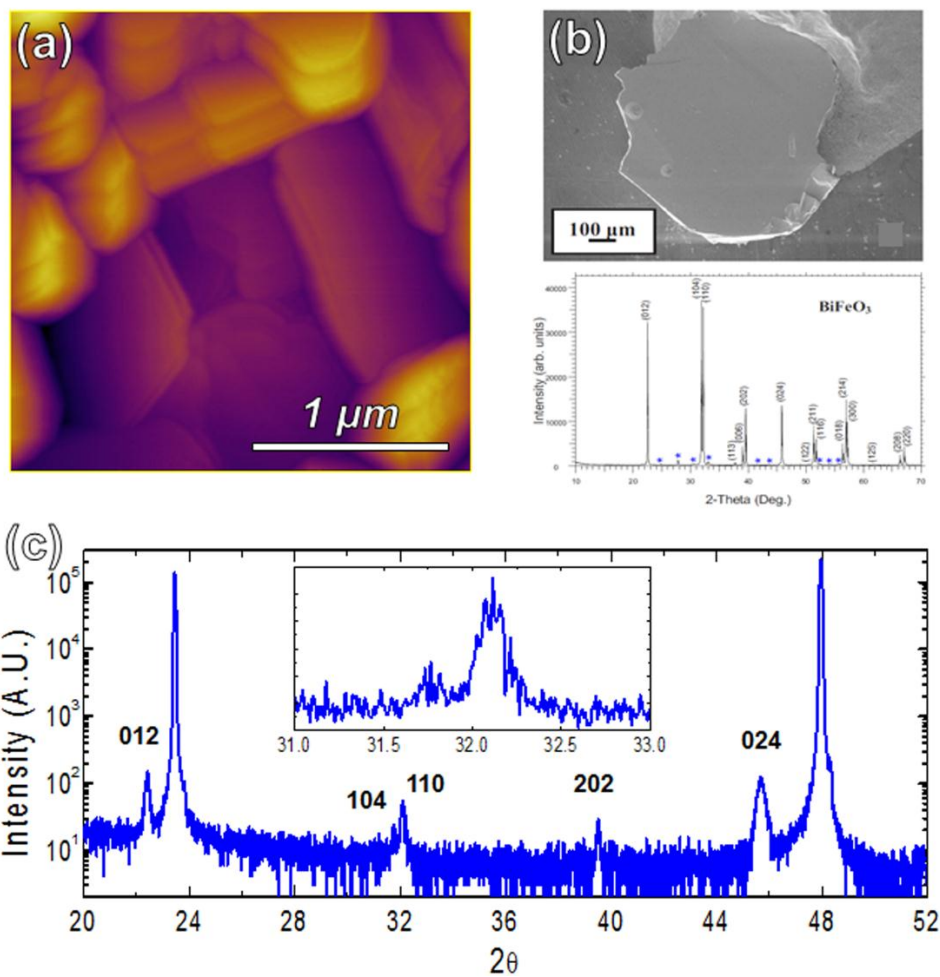

FIG. 3. (a) High-resolution atomic force microscopy image of micron-sized crystallites found in films > 250 nm thick. (b) Image and X-ray diffraction pattern of a flux grown rhombohedral BiFeO$_3$ sample. Figure courtesy of Ref. 26. (c) X-ray diffraction pattern of a 350 nm thick BiFeO$_3$/LaAlO$_3$ (001) heterostructure reveals signatures of the parent rhombohedral phase.

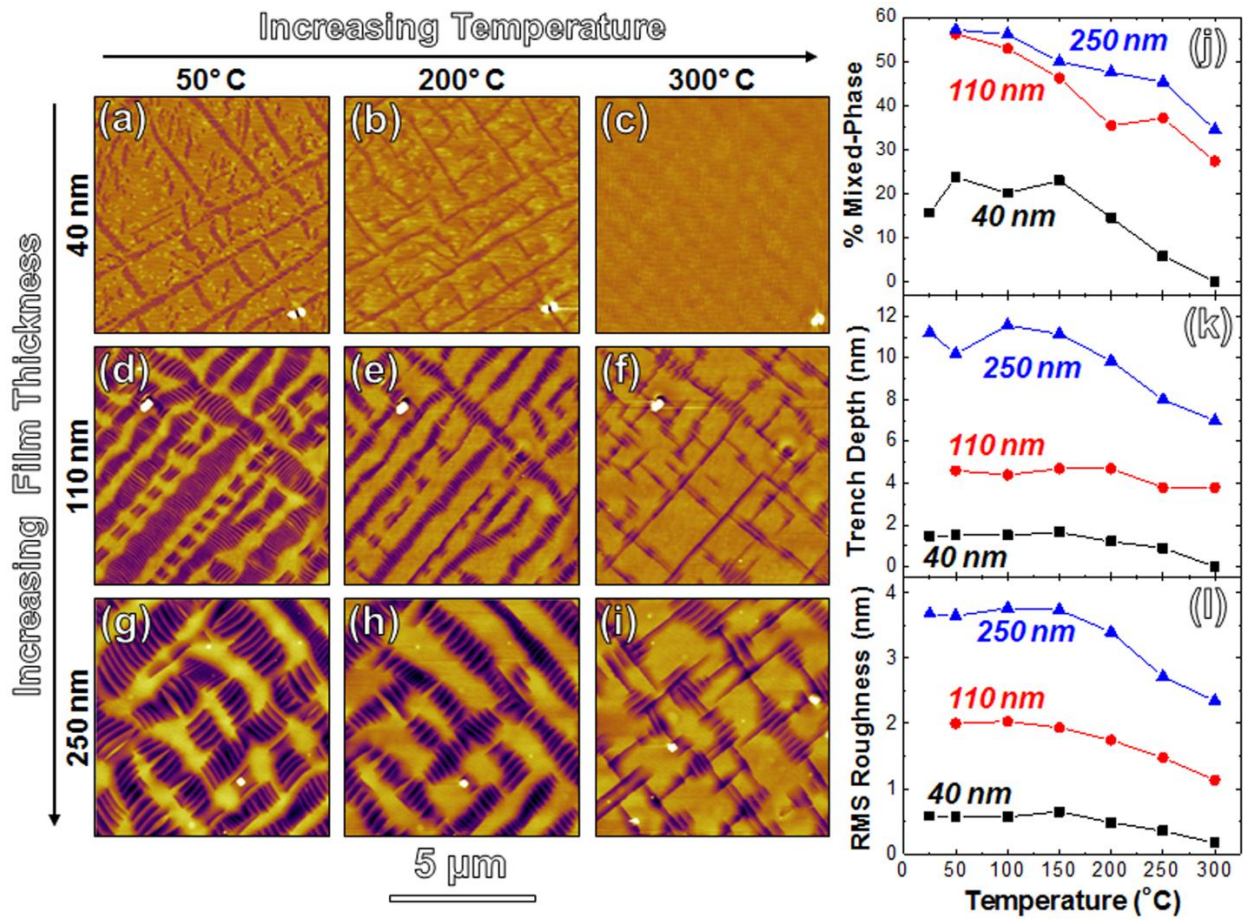

FIG. 4. Atomic force microscopy study of the evolution of surface morphology with increasing temperature from 50°C to 300°C for (a) – (c) 40 nm, (d) – (f) 110 nm, and (g) – (i) 250 nm thick films. Corresponding analysis of temperature-dependent evolution of properties including (j) the relative fraction of the mixed-phase structure at the surface, (k) the average depth of the mixed-phase trenches relative to the surrounding $M_{II}$-phase, and (l) the root-mean-square (RMS) roughness of the samples. Note the general trend to decrease the fraction of the mixed-phase region with increasing temperature and complete disappearance of the mixed-phase in thinner films.

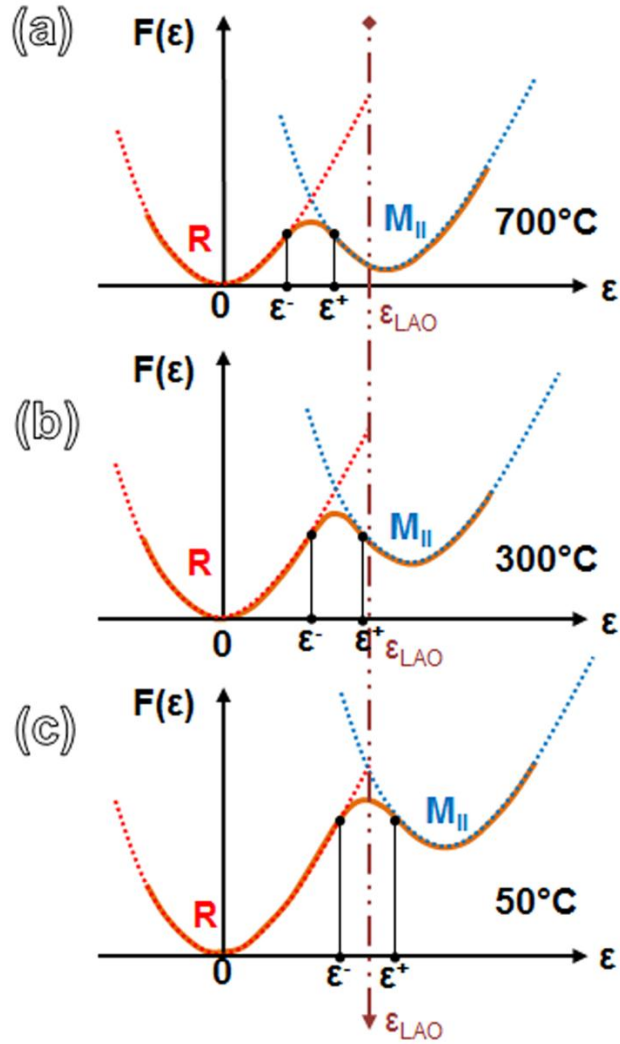

FIG. 5. Schematic illustration of the anticipated evolution of free energy of the system as a function of thin film strain. Upon transitioning from (a) 700°C to (b) 300°C to (c) 50°C we anticipate movement of the free energy curves such that spontaneous formation of the mixed-phase structures occurs as noted.

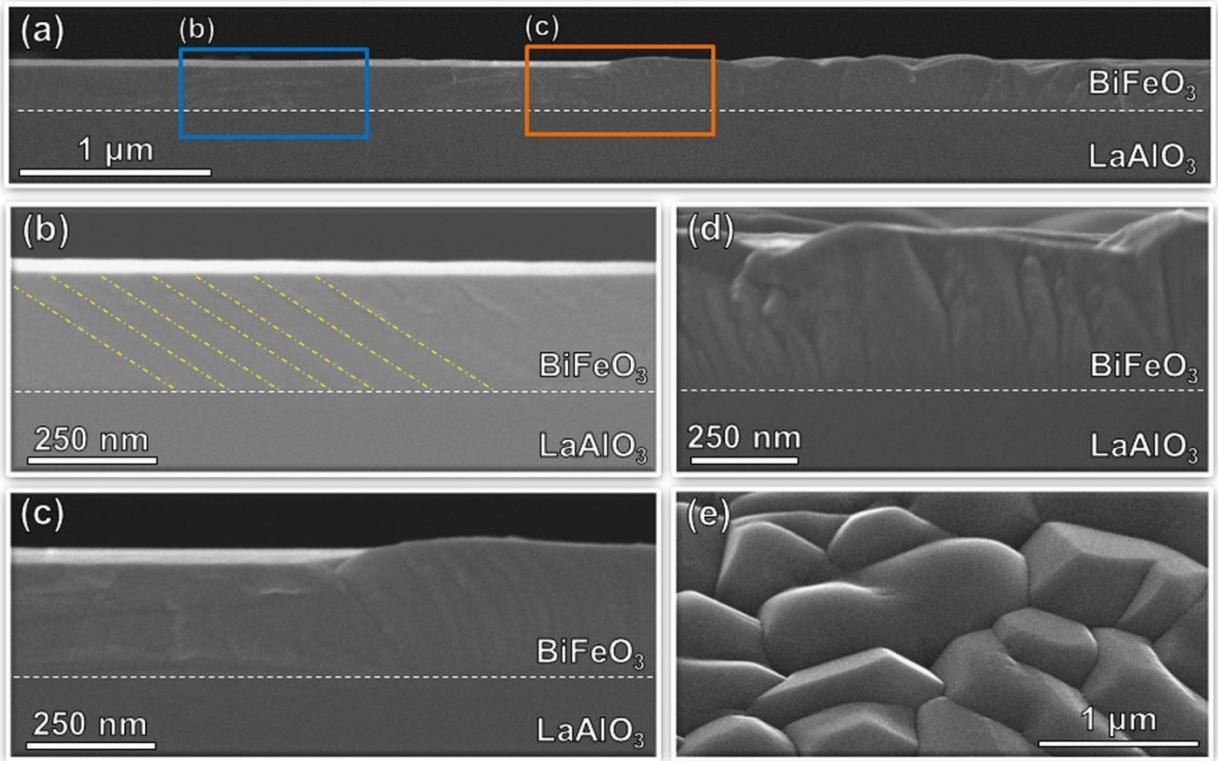

FIG. 6. Cross-sectional scanning electron microscope analysis of a 250 nm BiFeO$_3$ / LaAlO$_3$ (001) heterostructure. (a) Low-resolution view of sample shows transition from smooth to rough patches. (b) Close inspection of smooth areas reveals the presence of contrast consistent with mixed-phase region. (c) The rough, patchy regions are found to extend throughout the thickness of the film and have a fairly sharp boundary between regions. Analysis of thicker (350 nm) films reveals the presence of fully epitaxial breakdown with uniform structure throughout the thickness of the film (d) and the presence of faceted crystallites on the surfaces (e).

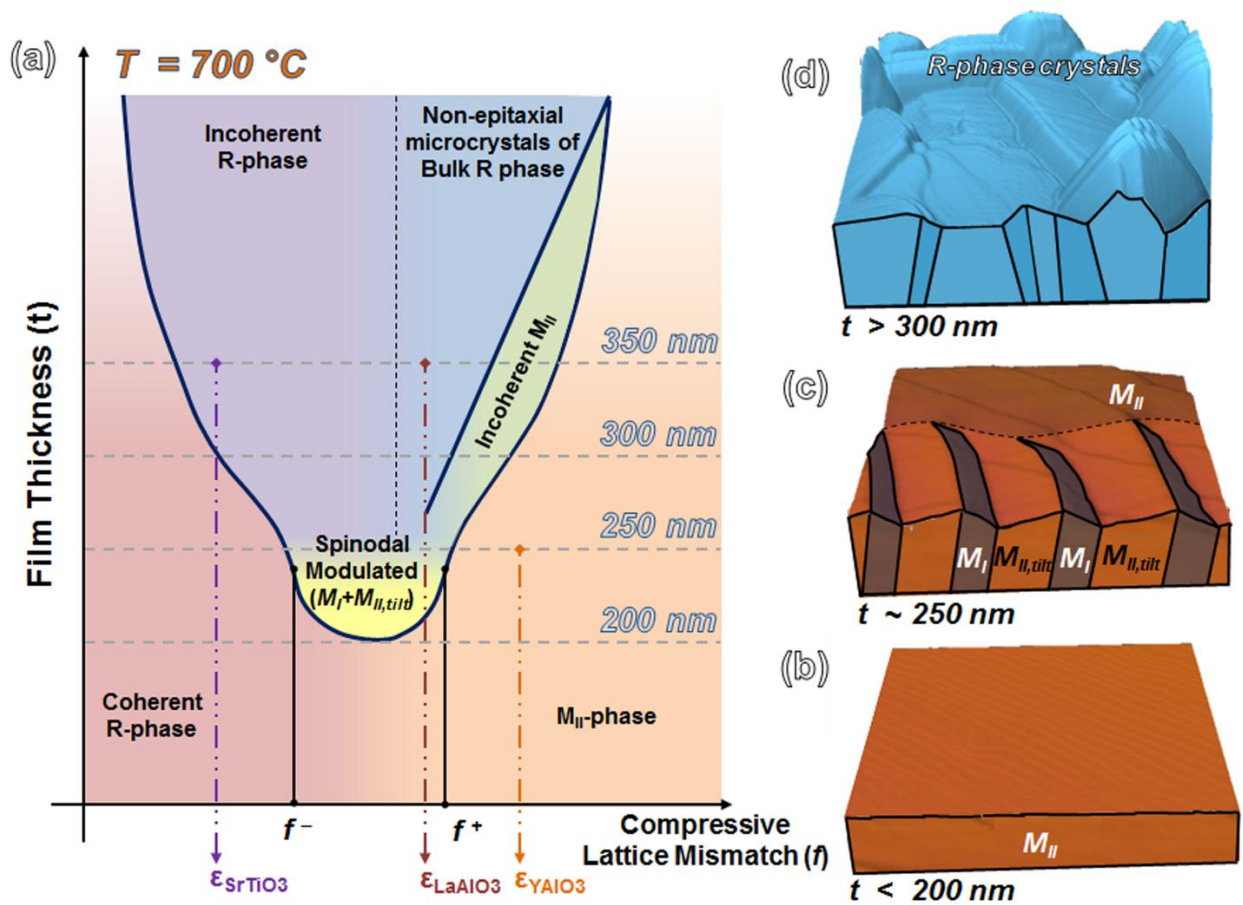

FIG. 7. (a) Schematic phase diagram showing the evolution of the microstructure as a function of epitaxial lattice mismatch (*f*) and film thickness (*h*). At the lattice mismatch expected between $BiFeO_3$ and $LaAlO_3$ we expect three different stages of growth: (b) coherent growth of the highly-distorted $M_{II}$-phase in thin films, (c) relaxation by formation of spinodal modulated structure of the $M_I$- and $M_{II,tilt}$-phase at intermediate thicknesses, and (d) eventual relaxation and transformation to non-epitaxial microcrystals of bulk R-phase.